\documentclass[prl,twocolumn,preprintnumbers,amsmath,amssymb]{revtex4}
\usepackage{graphicx}
\usepackage{dcolumn}
\usepackage{bm}
\usepackage{color}
\usepackage{braket}
\usepackage{mathtools}
\usepackage[english]{babel}
\usepackage{ulem}

\begin{document}

\title{Photoelectron Circular Dichroism of a Model Anionic System}

\author{\firstname{Anton~N.} \surname{Artemyev}}
\affiliation{Institut f\"{u}r Physik und CINSaT, Universit\"{a}t Kassel, Heinrich-Plett-Str. 40, 34132 Kassel, Germany}

\author{\firstname{Eric} \surname{Kutscher}}
\affiliation{Institut f\"{u}r Physik und CINSaT, Universit\"{a}t Kassel, Heinrich-Plett-Str. 40, 34132 Kassel, Germany}

\author{\firstname{Philipp~V.} \surname{Demekhin}}\email{demekhin@physik.uni-kassel.de}
\affiliation{Institut f\"{u}r Physik und CINSaT, Universit\"{a}t Kassel, Heinrich-Plett-Str. 40, 34132 Kassel, Germany}

\date{\today}

\begin{abstract}
Photoelectron circular dichroism in the one-photon detachment of a model methane-like chiral anionic system is studied theoretically by the single center method. The computed chiral asymmetry, characterized by the dichroic parameter $\beta_1$ of up to about $\pm3\%$, is in a qualitative agreement with very recent experimental observations on photodetachment in amino acid anions  [P. Kr\"uger and K.-M. Weitzel, Angew. Chem. Int. Ed. \textbf{60}, 17861 (2021)]. Our findings confirm a general  assumption that the magnitude of PECD is governed by the ability of an outgoing photoelectron wave packet to accumulate characteristic chiral asymmetry from the short-range part of the molecular potential. \end{abstract}

\maketitle

The photoelectron circular dichroism (PECD, \cite{Define}) is a chiroptical effect observed as a  forward-backward asymmetry in the emission of photoelectrons from randomly-oriented chiral molecules ionized by circularly-polarized light. Because PECD is a pure electric-dipole effect \cite{Ritchie}, it is much stronger than the conventional circular dichroism in the photoabsorption spectra of chiral molecules. This universal chiroptical effect emerges not only in one-photon ionization \cite{REV1,REV2,REV3}, as was initially predicted by Ritchie~\cite{Ritchie}, but in many other regimes, such as  resonance-enhanced multiphoton ionization \cite{Lux12AngChm,Lehmann13jcp}, above-threshold ionization  \cite{Beaulieu16NJP,Lux16ATI}, strong-field ionization \cite{Beaulieu16NJP,Fehre19}, and multiphoton ionization by bichromatic fields \cite{Beaulieu17as,Beaulieu18PXCD,PRLw2w,Rozen19}. The high contrast of the effect, seen in all photoionization regimes and in a broad range of photon energies, controllability by tailored light, and high  enenantiomeric excess resolution \cite{Kastner16ee} established PECD as a powerful tool for chiral recognition in the gas phase.

Ritchie~\cite{Ritchie} has shown theoretically that PECD emerges owing to an incomplete compensation between amplitudes for the population of partial photoelectron waves with positive and negative projections $\pm m$ of the carried angular momentum $\ell$. When escaping a molecule, the wave packet of an outgoing photoelectron accumulates this characteristic asymmetry by multiple scattering on the chiral molecular potential. It is commonly assumed that those effects are governed by the short-range interaction part of the molecular potential, and not from its long-range Coulomb interaction part necessarily present in an ion. However, it is rather impossible to separate individual contributions of the short- and long-range interaction parts of an ionic potential in any ab-initio calculations of a photoionization process. Here, a natural way to exclude the long-range Coulomb part of the potential is to study photodetachment of anions \cite{ADanions}.

Anions (negative ions with more electrons than protons) are very common in nature \cite{REVanion1,REVanion0} with a great abundance in the biological world \cite{REVanion2}. Importantly, almost all biomolecules are chiral, and their enantiomers often exhibit different bioactivity \cite{drags}.  In addition, anions have rather low detachment energies \cite{REVanion1,REVanion0}, typically within the reach of a single UV photon, which nowadays can be provided in table-top experiments. It is, therefore, rather natural to exploit PECD for recognition of chiral anions, as is proposed in a very recent work~\cite{EXPTanions} and demonstrated by measuring a sizable chiral asymmetry in photodetachment of chiral amino acid anions. Experiments, performed in Ref.~\cite{EXPTanions} with different enantiomers of 3,4-dihydroxyphenylalanine and glutamic acid, confirmed that short-range interaction potentials of these chiral anions are sufficient to generate chiral asymmetries PECD$=2\beta_1$ of about $\pm 4$\%.

The gas-phase laboratory-frame angular distribution of photoelectrons emitted from molecules, which are ionized by the absorption of one circularly-polarized photon, is given by the total cross section $\sigma$, dichroic parameter $\beta_1$, and anisotropy parameter $\beta_2$ via \cite{TFMOX,Hartmann19}:
\begin{equation}
\frac{\mathrm{d}\sigma^\pm(\theta)}{\mathrm{d\Omega}}=\frac{\sigma}{4\pi} \left[ 1 \pm \beta_1 P_1(\cos \theta)  -\frac{1}{2}\beta_2  P_2(\cos \theta)\right]. \label{eq:DPICS}
\end{equation}
Here, `$\pm$' stand for the positive or negative helicity of the light, $P_L$ are the Legendre polynomials, and $\theta$ is the angle between the direction of the propagation of the ionizing radiation and the direction of the emission of photoelectrons. There are a few approaches to model laboratory-frame angular distributions in photodetachment of anions \cite{ADanions}. Those models, however, allow for interpretation of the anisotropy parameter $\beta_2$, which is an additive quantity given by a coherent superposition of all partial photoionization amplitudes. Accurate description of the dichroic parameter $\beta_1$, which is a subtractive quantity depending on differences between partial amplitudes (see above and Ref.~\cite{Ritchie}) and which is very sensitive to each and every particular detail of the  molecular potential, requires reliable description of the electron continuum spectrum in molecules.

\begin{figure}
\includegraphics[width=0.465\textwidth]{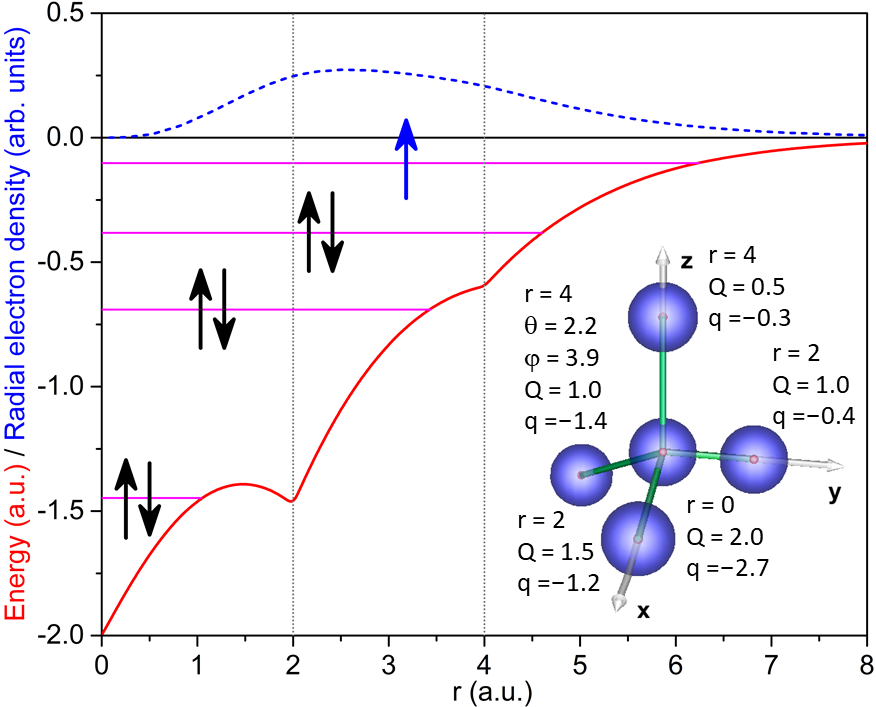}
\caption{Inset: model methane-like neutral chiral system, which is built of 5 point charges $Q$ (red dots) located at positions $(r,\theta,\varphi)$ and surrounded by spherically symmetric electron densities (blue clouds) with the total charges $q$ and charge distributions $e^{-r}$. For transparency,  positions and sizes of the constituent parts are depicted without keeping proportions. All parameters are given in a.u. The total nuclear charge of the system is +6, and the total electron charge is --6. Red curve: spherical part of this neutral chiral potential well. Vertical grey dotted lines indicate positions of nuclei at 2 and 4~a.u. The energy spectrum of the bound states, supported by this well, is indicated by the horizontal solid magenta lines. The three lowermost states are assumed to be occupied by six electrons. Introducing an excess electron to the fourth level creates an anion. Blue dashed curve: radial electron density of the excess electron, involved in photodetachment from the fourth bound state.}  \label{fig1:model}
\end{figure}

Although there are a few well-established theoretical approaches \cite{MSXalp,TDDFT,SC1,SC2} to study one-photon PECD in photoionization of neutral chiral molecules, there are, to the best of our knowledge, no theoretical studies of PECD in photodetachment of chiral anions. In the present work, we close this gap. To this end, we model the electronic potential of a chiral anionic system and apply a reliable theoretical approach to study its angle-resolved photodetachment by one-photon absorption. In particular, we make use of a model methane-like chiral system which was introduced in our previous work \cite{TDSC1}. It consists of five point charges each screened by surrounding spherically symmetric electron densities. A sketch of the system and all parameters used in the calculations are indicated in the inset in Fig.~\ref{fig1:model}. These parameters were chosen to generate a strongly asymmetric chiral one-particle potential for the initial state and for the photoelectron. In addition, the total positive charge of all nuclei (+6) is fully compensated by the total negative charge of the electron clouds (--6), which simulates an overall neutral potential for an additional electron.

The spherical part of this potential is depicted in Fig.~\ref{fig1:model} by the  red solid curve. The energy spectrum of the first four bound electronic states, supported by the presently generated three-dimensional one-particle chiral potential well, is indicated in Fig.~\ref{fig1:model} by the magenta horizontal lines. The three lowermost energy levels can be associated with the doubly-occupied  core and valence orbitals of the neutral system. Populating the fourth bound state of the potential by a single electron produces, thus, an anion. This fourth bound state (with the binding energy of 2.78~eV) was chosen as the initial electronic state to study the photodetachment process. The radial electron density of this initial state (with a mean radius of about 3.4~a.u.) is depicted in the upper part of Fig.~\ref{fig1:model} by the blue dotted curve.

\begin{figure}
\includegraphics[width=0.465\textwidth]{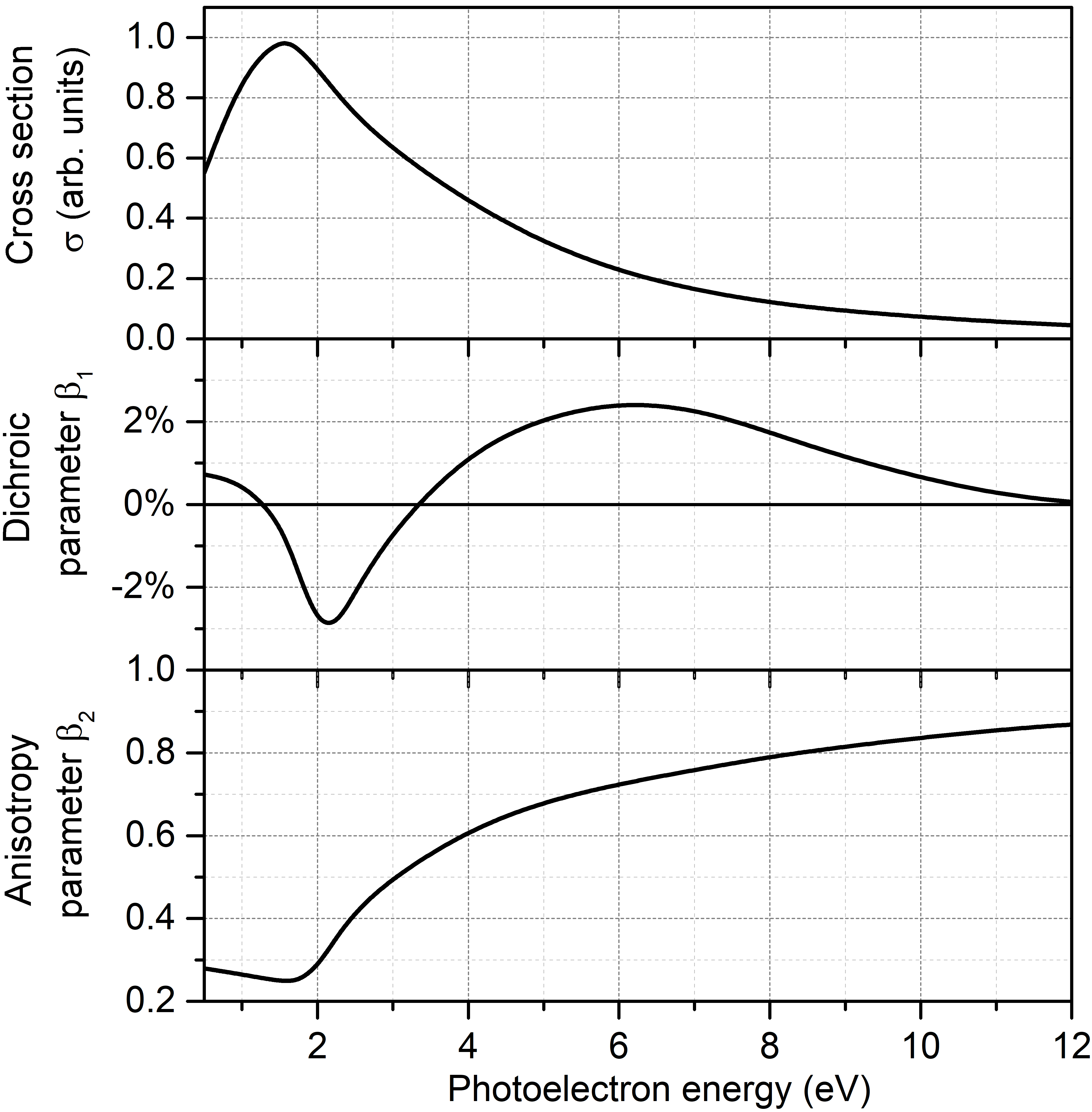}
\caption{Cross section (upper panel), dichroic parameter (middle panel) and anisotropy parameter (lower panel) as functions of the photoelectron kinetic energy, computed for one-photon detachment of the model methane-like chiral anionic system from Fig.~\ref{fig1:model}.}  \label{fig2:calculations}
\end{figure}

The present calculations were performed by the time-dependent single center method and code \cite{TDSC1,TDSC2,TDSC3}. In particular, we numerically propagated wave packets of the excess electron in the field of the neutral chiral potential well. Those wave packets were  released by weak Gaussian-shaped circularly-polarized pulses with the carrier frequencies above the photodetachment threshold of 2.78~eV (i.e., photodetachment by a one-photon absorption). Details on the calculations can be found in Ref.~\cite{TDSC1} and only two essential points relevant for the present work are mentioned below. In order to obtain final (after the end of the pulses) electron momentum distribution in the anion, the spatial photoelectron wave packets were projected on the plane waves (i.e., Fourier-transformed) and not on the Coulomb waves, which are usually used to analyze photoionization of neutrals. The laboratory-frame angle-resolved photodetachment spectra were obtained by numerical integration over molecular orientation Euler angles in steps of $\Delta \alpha= \Delta \beta=0.1\,\pi$. This numerical averaging procedure was cross-checked using analytically averaged  expressions for the dichroic $\beta_1$ and anisotropy $\beta_2$ parameters \cite{TFMOX}.

Results of the present calculations are summarized in Fig.~\ref{fig2:calculations}. As one can see from the uppermost panel of this figure, the total one-photon detachment cross section exhibits a typical shape resonance just above the threshold \cite{TFMOX} (here, at the photoelectron kinetic energy of about 1.5~eV). Its signature is also clearly seen in the anisotropy parameter $\beta_2$ around the photoelectron energy of 2~eV (lowermost panel of the Fig.~\ref{fig2:calculations}). This shape resonance gives also rise to a very broad feature \cite{TFMOX} in the computed dichroic parameter $\beta_1$ (middle panel of the figure) and causes change of its sign twice across the resonance \cite{Hartmann19}. As expected \cite{Stener04,Powis00}, the computed dichroic parameter vanishes for sufficiently large  photoelectron kinetic energies (here, larger than about 12~eV). Importantly, the magnitude of the dichroic parameter (in between $\pm3\%$), computed for the one-photon detachment of this model anionic system, is rather typical for a one-photon ionization of neutral chiral molecules \cite{REV1,REV2,REV3}.

In conclusion, we theoretically demonstrate a sizable PECD in the one-photon detachment of a model methane-like chiral anionic system. Our numerical findings qualitatively support the recent idea  \cite{EXPTanions} to extend PECD as a  gas-phase tool for recognition of chiral anions.  The fact that the short-range potential is capable to imprint molecular chirality on the outgoing photoelectron wave packet, is also important for chiral recognition in the nanoparticle \cite{nanoP} and liquid \cite{Liquid} phases, where most of the long-range Coulomb interaction is shielded.

\section{acknowledgements}
The authors acknowledge Uwe Hergenhahn for motivation to perform this work and for many valuable discussions, including liquid-phase perspectives of the present results. This work was funded by the Deutsche Forschungsgemeinschaft (DFG) -- Project No. 328961117 -- SFB 1319 ELCH (Extreme light for sensing and driving molecular chirality, project C1).

\section{AUTHOR DECLARATIONS}

\subsection{Conflict of Interest}
The authors have no conflicts to disclose.

\subsection{Author Contributions}
The authors have contributed equally to this work.

\section{DATA AVAILABILITY}
The data that support the findings of this study are available from the corresponding author upon reasonable request.

\end{document}